\documentclass[prd,twocolumn,showpacs,preprintnumbers,superscriptaddress]{revtex4}
\usepackage{hyperref,amssymb,amsmath,mathrsfs,bm,graphicx}
\begin{document}
\title{Heat flow in the postquasistatic approximation}
\author{B. Rodr\'\i guez--Mueller}
\affiliation{Computational Science Research Center, College of Sciences,
San Diego State University, San Diego, California, USA}
\author{C. Peralta}
\affiliation{Deutscher Wetterdienst, Frankfurter Str. 135, 63067 Offenbach, Germany}
\affiliation{School of Physics, University of Melbourne, Parkville,
VIC 3010, Australia}
\author{ W. Barreto}
\affiliation{Centro de F\'\i sica Fundamental, Facultad de Ciencias, Universidad de Los Andes, M\'erida, Venezuela\footnote{On sabbatical leave.}} 
\author{L. Rosales}
\affiliation{Laboratorio de F\'\i sica Computacional,
Universidad Experimental Polit\'ecnica ``Antonio Jos\'e de
Sucre'', Puerto Ordaz, Venezuela}
\date{\today}

\begin{abstract}
We apply the postquasistatic approximation to study the evolution
of spherically symmetric fluid distributions undergoing dissipation
in the form of radial heat flow.   
For a model which corresponds to an incompressible fluid departing from the static equilibrium,
it is not possible to go far from the initial state after the emission of a
small amount of energy. Initially
collapsing distributions of matter are not permitted. Emission
of energy can be considered as a mechanism to avoid the collapse. 
If the distribution collapses initially and emits one hundredth of the initial
mass only the outermost layers evolve.
For a model which corresponds to a highly compressed Fermi gas, only the outermost 
shell can evolve with a shorter hydrodynamic time scale.
\end{abstract}
\maketitle
\section{Introduction}
Dissipation as an emission process \cite{ks79} is crucial
for the outcome of gravitational collapse. 
Thermal conduction is usually considered proportional to the gradient of temperature. This
is a sensible choice, since the mean free path of particles responsible for the propagation 
of energy in stellar interiors is very small as compared with the typical length of the object \cite{hd97}.
Observations from supernova 1987A indicate that the regime of radiation transport prevailing during the emission process is closer to the diffusion approximation than to the free streaming limit \cite{l88}. The addition of a test bed for studying dissipation mechanisms and
other transport processes in order to later incorporate them into a 
more sophisticated numerical framework (Arnowitt--Deser--Misner [ADM] 
or characteristic) is a necessity.

In this work we study a self--gravitating spherical distribution of matter containing a dissipative fluid in the diffusion limit.
We found behaviors similar to those reported with a
different mechanism by \cite{prrb10}, and report the zeroth order results for
dissipation.
We use noncomoving coordinates and follow the method reported in \cite{brm02}, \cite{hbds02} named the postquasistatic approximation (PQSA), which has been proposed
as a test bed in numerical relativity \cite{b09}.
For recent advances and applications see  \cite{bcb09} and \cite{bcb10}.
For origin, reviews and details of the PQSA see \cite{hjr80}, \cite{almostall1}--\cite{almostall7} and \cite{allheatflow1}--\cite{allheatflow8}.
We do not consider here temperature profiles to determine which processes can take place during the collapse. For that purpose, transport equations in the relaxation time approximation have been proposed to avoid pathological behaviors (see for instance \cite{m96} and references therein). These issues will be considered in a future investigation. It is worth mentioning here that in order to get a higher order approximation we have to know the zero order approximation in the relaxation time, as in the present study.
To the best of our knowledge, no author has undertaken in practice the dissipative matter problem in numerical relativity. Our purpose here is to show how heat flow processes can be considered
in the context of the PQSA. The results indicate that an observer
using radiation coordinates does not ``see" some details when heat flow is considered. The final goal is to eventually study the same problem using the M\"uller--Israel--Stewart theory for the dissipative system, which is highly nontrivial in spherical symmetry.

In Sec. II, we present the field equations, the matching conditions and the set of surface equations.
For additional details concerning the PQSA method see \cite{hbds02} and \cite{prrb10}. Three models are presented in Sec. III and some remarks are
discussed in Sec. IV.
  
\section{Main equations}
~~~~To write the Einstein field equations we use the line element in Schwarzschild--like coordinates
\begin{equation}
ds^2=e^\nu dt^2-e^\lambda dr^2-r^2\left( d\theta ^2+\sin
{}^2\theta d\phi ^2\right), \label{eq:metric} 
\end{equation}
where $\nu = \nu(t,r)$ and $\lambda = \lambda(t,r)$, with
 $(t,r,\theta,\phi)\equiv(0,1,2,3)$.

In order to get physical input we introduce the
 Minkowski coordinates $(\tau,x,y,z)$ by \cite{b64}
\begin{equation}
d\tau=e^{\nu /2}dt,\,  
dx=e^{\lambda /2}dr,\,  
dy=rd\theta,\, 
dz=r \sin \theta d\phi,\label{eq:local}
\end{equation}
In these expressions $\nu$ and $\lambda$ are constants, because they
 have only local values.
 
Following the Bondian point of view as in \cite{b64}, \cite{hbds02}, 
\cite{b09} and \cite{prrb10}
we assume that, for an observer moving relative to the local Minkowskian coordinates
 with velocity $\omega$ in the radial direction, the space contains
 an isotropic fluid of energy density $\rho$, radial pressure $p$, and
radial heat flux $q$.
For this comoving observer, the covariant energy tensor in Minkowski
 coordinates is thus
 
\begin{equation}
\left(
\begin{array}{cccc}
\rho & -q & 0 & 0 \\
-q & p & 0 & 0 \\
0 & 0 & p & 0 \\
0 & 0 & 0 & p 
\end{array}
\right),
\end{equation}
Making a Lorentz boost we write the field equations in relativistic units ($G=c=1$)
as follows \cite{prrb10}:
\begin{equation}
\tilde\rho =
\frac{1}{8\pi r}\left[\frac{1}{r} - 
e^{-\lambda}\left(\frac 1{r}-\lambda_{,r}\right)\right], \label{eq:ee1}
\end{equation}

\begin{equation}
\tilde p =
\frac{1}{8\pi r}\left[
e^{-\lambda}\left(\frac 1{r}+\nu_{,r}\right) - \frac{1}{r}\right], \label{eq:ee2}
\end{equation}

\begin{eqnarray}
p= &&\frac{1}{32\pi} \{ e^{-\lambda}[ 2\nu_{,rr}+\nu_{,r}^2
-\lambda_{,r}\nu_{,r} + \frac{2}{r}
(\nu_{,r}-\lambda_{,r}) ] \nonumber \\ \nonumber \\
&-&e^{-\nu}[ 2\lambda _{,tt}+\lambda_{,t}(\lambda_{,t}-\nu_{,t}) ] \}, \label{eq:ee3}
\end{eqnarray}

\begin{equation}
S = 
-\frac{\lambda_{,t}}{8\pi r}e^{-\frac 12(\nu+\lambda)}, \label{eq:ee4}
\end{equation}
where the comma (,) represents partial differentiation with 
 respect to the indicated
 coordinate and
the conservative variables are
 \begin{equation}
\tilde\rho= \frac{\rho + p \omega^2}{1-\omega ^2} + \frac{2 q \omega}{1 - \omega^2}, \label{eq:ev1}
\end{equation}
\begin{equation}
S=(\rho + p)\frac{\omega}{1-\omega^2} + q\frac{ 1+\omega^2}{1-\omega^2}
\end{equation}
and the flux variable
\begin{equation}
\tilde p = \frac{p + \rho \omega^2}{1-\omega ^2} + \frac{2 q \omega}{1 - \omega^2}. \label{eq:ev2}
\end{equation}
as in the standard ADM 3+1 formulation. Within the PQSA $\tilde\rho$ and $\tilde p$ are
referred as to effective density and effective pressure, respectively.
Note that from (\ref{eq:local}) the velocity of matter in Schwarzschild
 coordinates is
\begin{equation}
\frac{dr}{dt} = \omega e^{(\nu-\lambda)/2}. \label{eq:velocity}
\end{equation}

It is easy to check that \cite{hd97} 
\begin{equation}
 p_a = q_a, \label{eq:boundary}
\end{equation}
which expresses the continuity of the radial pressure across the boundary of the
distribution $r=a(t)$.
Equivalently, in terms of the effective variables
 \begin{equation}
\tilde p_a=\tilde\rho_a\omega_a^2 + q_a (1 + \omega_a)^2.
 \label{eq:boun}
\end{equation}

Defining the mass function as
\begin{equation}
e^{-\lambda}=1-2m/r, \label{eq:mass}
\end{equation}
and substituting (\ref{eq:mass}) into (\ref{eq:ee1}) and (\ref{eq:ee4})
 we obtain, after some rearrangements, 
 \begin{equation}
\frac{dm}{dt}=-4\pi r^2\left[\frac{dr}{dt}p
+q (1-2m/r)^{1/2}e^{\nu/2} \right]. \label{eq:energy}
\end{equation}
This equation is the momentum constraint in the ADM 3+1 formulation, it expresses
the power across any moving spherical shell.

Equation (\ref{eq:ee3}) can  be written as $T_{1;\mu }^\mu=0$ or equivalently,
after a lenghty calculation
\begin{eqnarray}
&&\tilde p_{,r} + \frac{(\tilde\rho + \tilde p)(4\pi r^3\tilde p + m)}{r(r-2m)}
 + \frac{2}{r}(\tilde p - p)\nonumber \\
&=&\frac{e^{-\nu}}{4\pi r(r-2m)}\left( m_{,tt} +\frac{3m_{,t}^2}{r-2m}-
\frac{m_{,t}\nu_{,t}}{2}\right) .\label{eq:TOV}
\end{eqnarray}
This last equation is the generalization of the Tolman--Oppenheimer--Volkov
 for nonstatic radiative situations. In can be shown that Eq. (\ref{eq:TOV})
 is equivalent to the equation of motion for the fluid in conservative form
 in the standard ADM 3+1 formulation \cite{b09}.
 
 At the surface, Eqs. (\ref{eq:velocity}), (\ref{eq:energy}) and (\ref{eq:TOV}) 
 lead us to a set of differential equations for $a$, $m_a$ and $\omega_a$ if we prescribe in some
 way the metric functions ($m$ and $\nu$).
 
 The other two field equations (\ref{eq:ee1}) and (\ref{eq:ee2}) can be integrated to obtain
 \begin{equation}
m= \int^{r}_{0}4\pi r^2 \tilde \rho \ dr \label{eq:m}
\end{equation}
which is the Hamiltonian constraint in the ADM 3+1 formulation and
\begin{equation}
\nu=\nu_{a} + \int^r_a \frac{2(4\pi r^3 \tilde p + m)}{r(r-2m)}dr, \label{eq:nu}
\end{equation}
the polar slicing condition, from where it is obvious that for a given radial dependence of the effective
variables, the radial dependence of the metric functions becomes completely determined.

\begin{figure}[htbp!]
\begin{center}
\scalebox{0.35}{\includegraphics[angle=0]{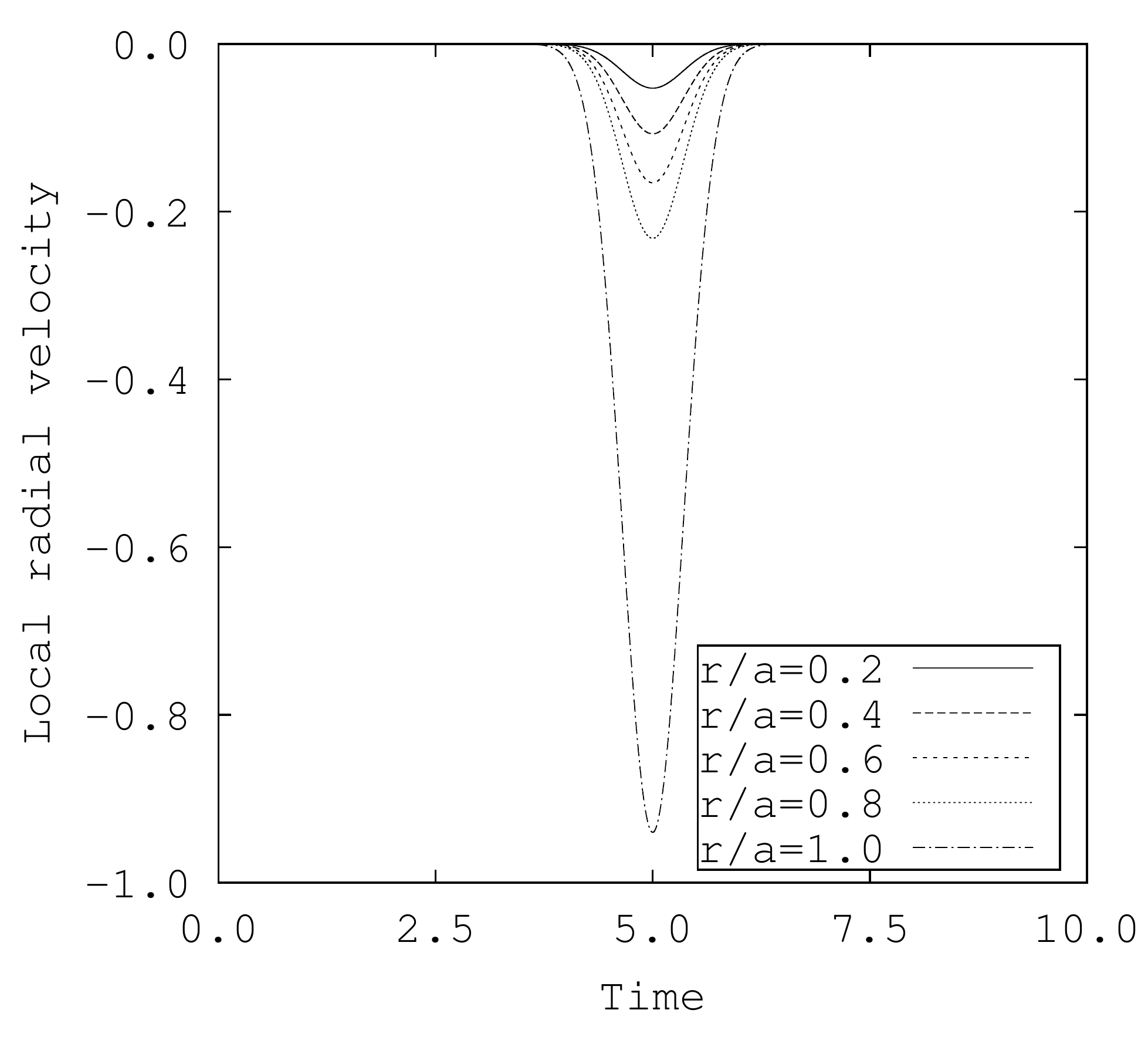}}
\caption{Evolution of the radial local velocity $\omega$ (multiplied by $10^3$) for the Schwarzschild--like model. 
The initial conditions are $a(0) = 5.0$, $m(0)=1.0$, $\omega_a(0)=0.0$.
The total radiated mass is $M_r= 10^{-4}\,m_a(0)$.}
\end{center}
\label{fig:omegar}
\end{figure}

\begin{figure}[htbp!]
\begin{center}
\scalebox{0.35}{\includegraphics[angle=0]{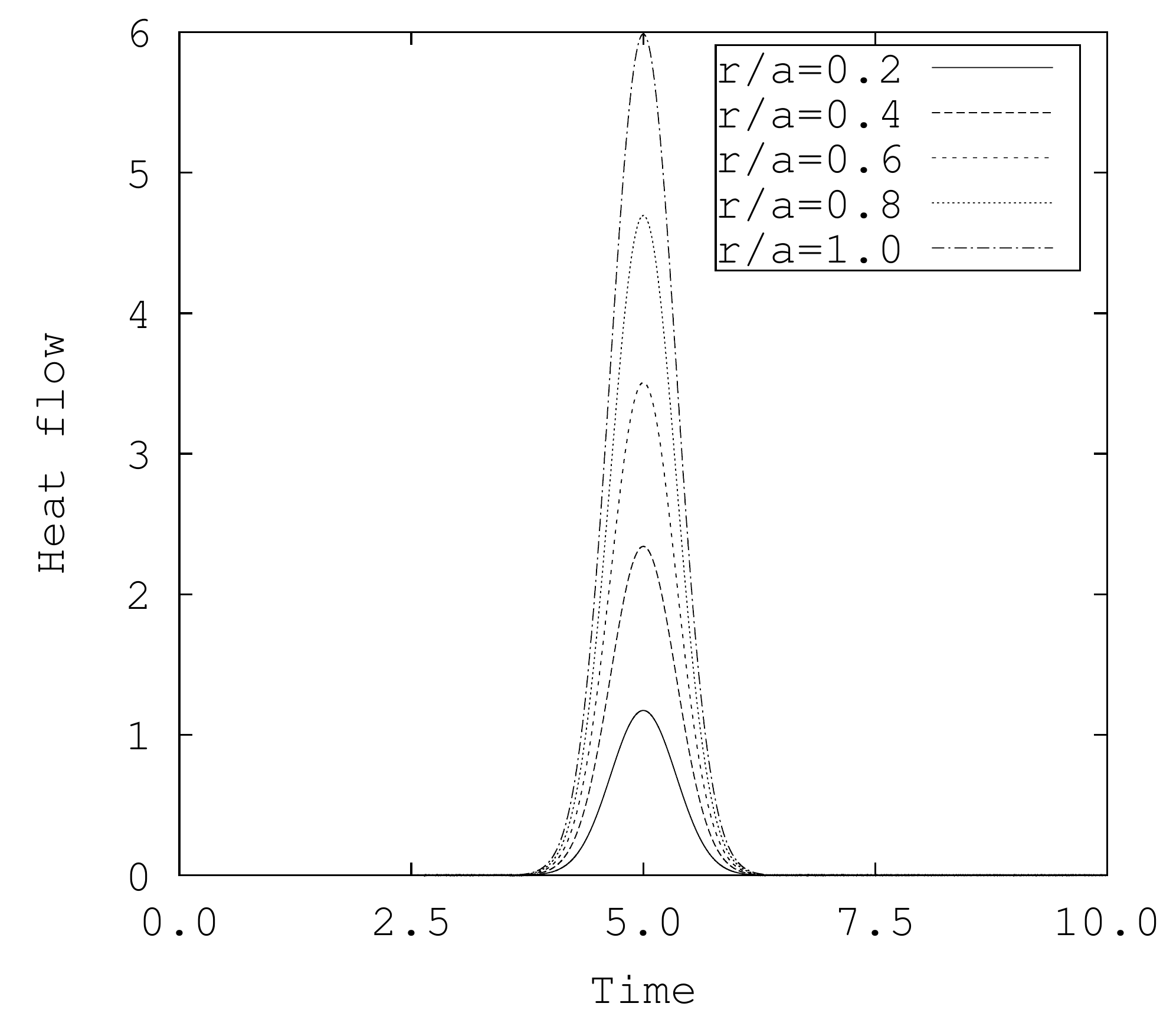}}
\caption{Evolution of the heat flow $q$ (multiplied by $10^7$) for the Schwarzschild--like model. The initial conditions are $a(0) = 5.0$, $m(0)=1.0$, $\omega_a(0)=0.0$.
The total radiated mass is $M_r= 10^{-4}\,m_a(0)$.}
\end{center}
\label{fig:flux}
\end{figure}

\begin{figure}[htbp!]
\begin{center}
\scalebox{0.35}{\includegraphics[angle=0]{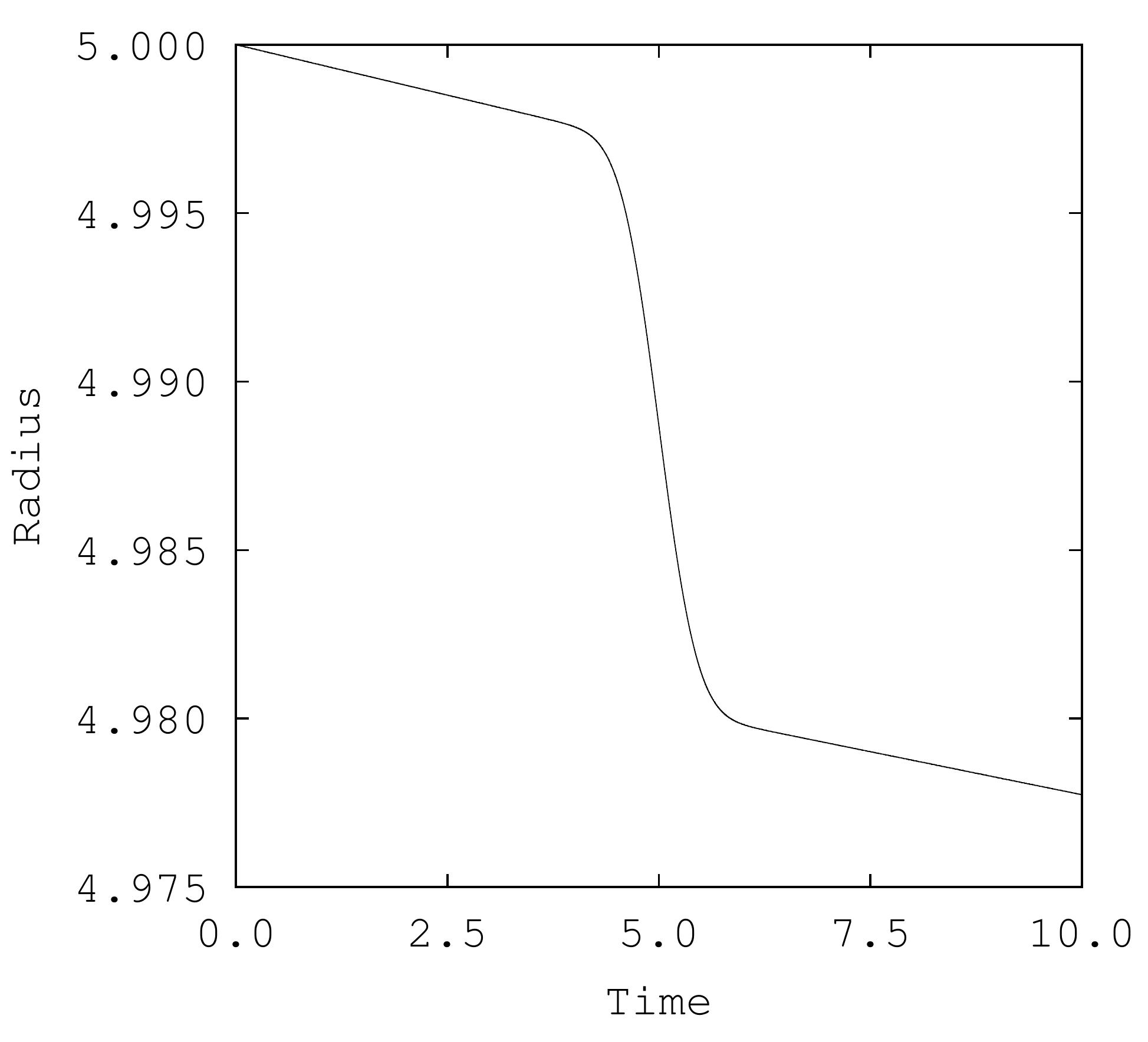}}
\caption{Evolution of the radius $a$ for the Schwarzschild--like model. The initial conditions are $a(0) = 5.0$, $m(0)=1.0$, $\omega_a(0)=-0.001$.
The total radiated mass is $M_r= 10^{-2}\,m_a(0)$.}
\end{center}
\label{fig:radius}
\end{figure}

\begin{figure}[htbp!]
\begin{center}
\scalebox{0.35}{\includegraphics[angle=0]{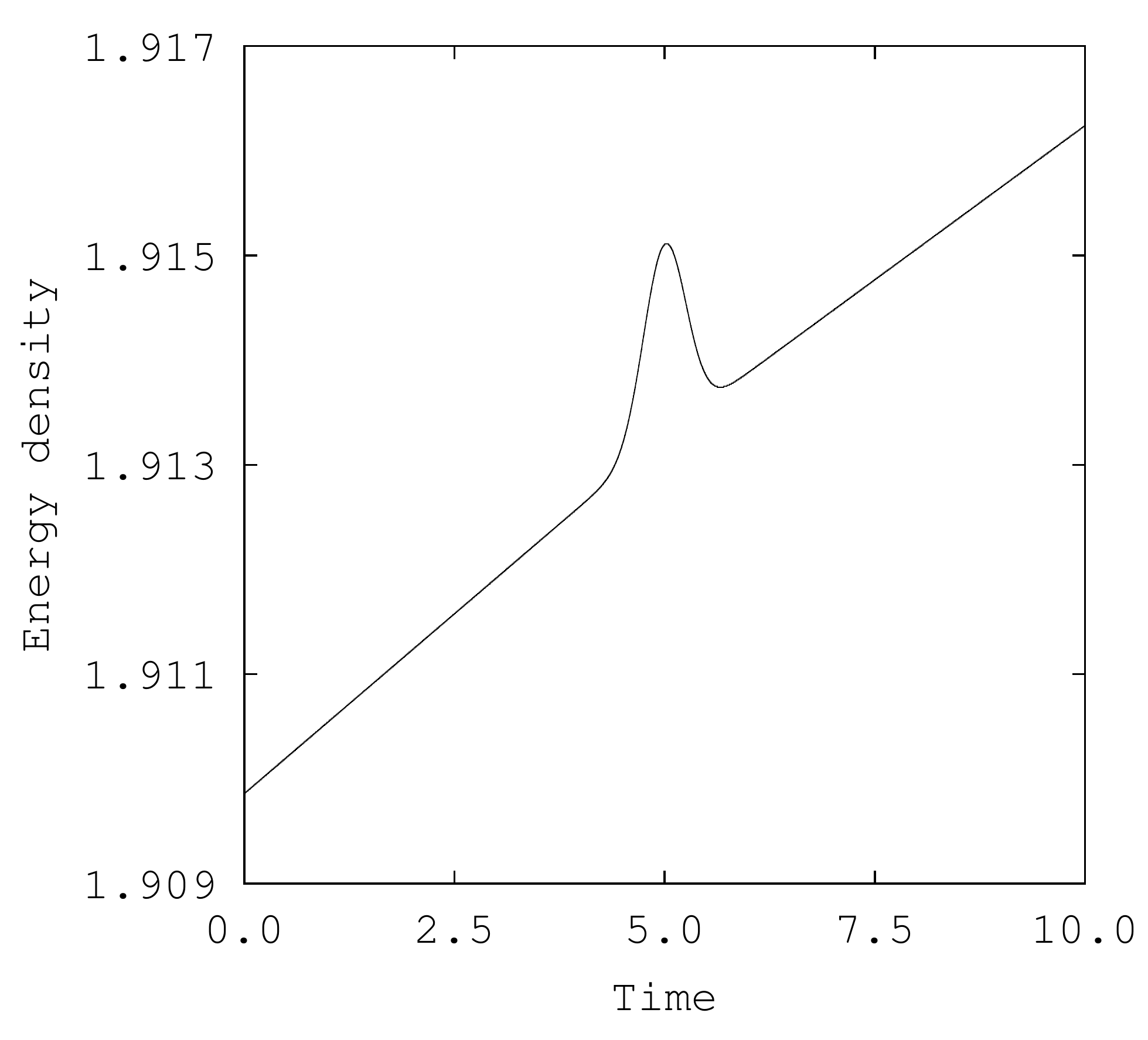}}
\caption{Evolution of the energy density $\rho$ (multiplied by $10^3$) for the Schwarzschild--like model. The initial conditions are $a(0) = 5.0$, $m(0)=1.0$, $\omega_a(0)=-0.001$.
The total radiated mass is $M_r= 10^{-2}\,m_a(0)$.}
\end{center}
\label{fig:dens}
\end{figure}

\begin{figure}[htbp!]
\begin{center}
\scalebox{0.35}{\includegraphics[angle=0]{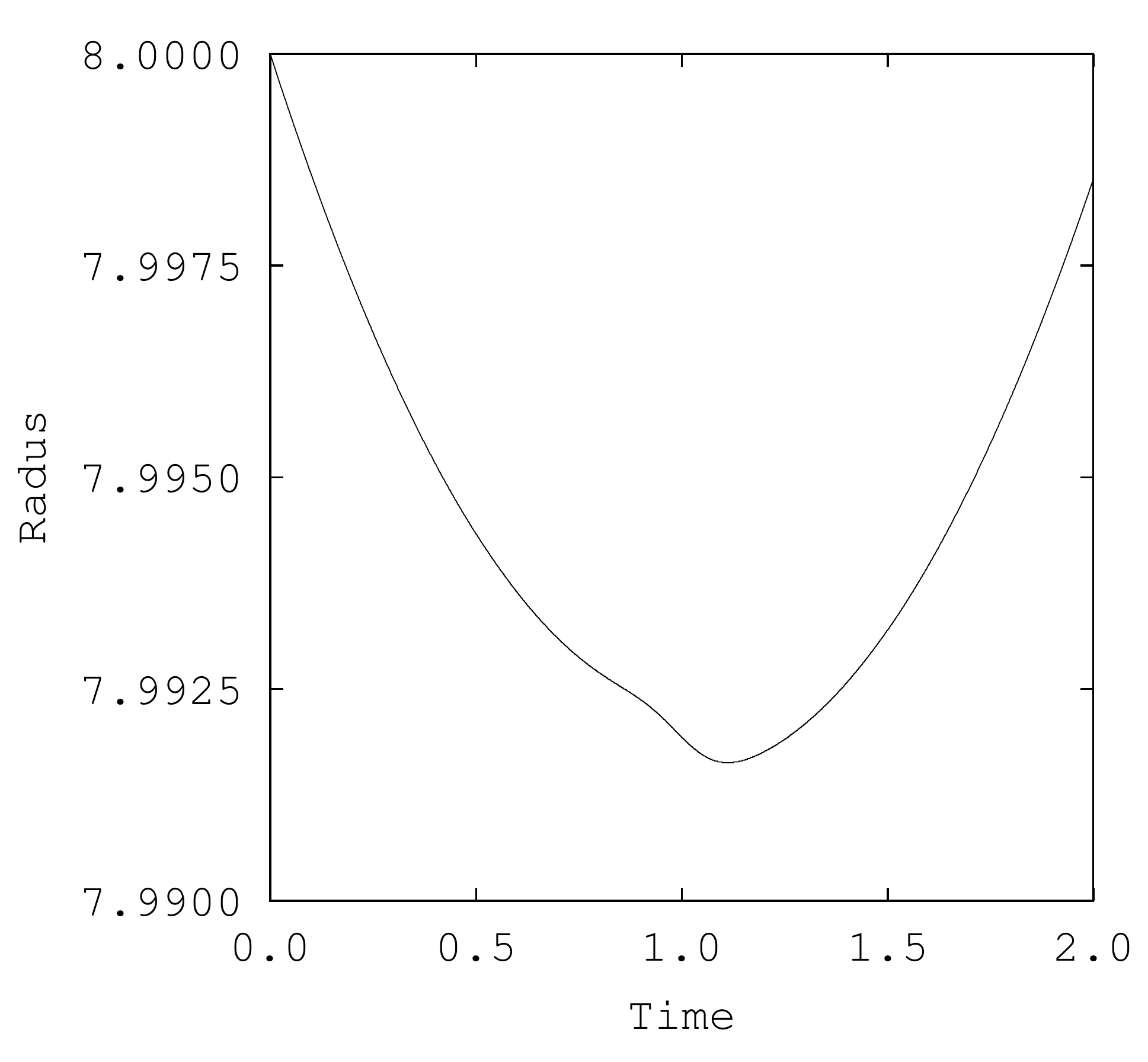}}
\caption{Evolution of the radius $a$ for the Tolman VI--like model. The initial conditions are $a(0) = 8.0$, $m(0)=1.0$, $\omega_a(0)=-0.02$.
The total radiated mass is $M_r= 10^{-4}\,m_a(0)$, with a narrow Gaussian given by $\Sigma=0.01$ with maximum at $t_0=1.0$.}
\end{center}
\label{fig:radio}
\end{figure}

\section{Modeling}
We consider here a seed model inspired by the well--known
Schwarzschild interior solution. This  model corresponds to an incompressible fluid departing from the static equilibrium. Following the PQSA we take
\begin{equation}
\tilde \rho =f(t),
\end{equation}
where $f$ is an arbitrary function of $t$.
The expression for $\tilde p$ is
\begin{equation}
\frac{\tilde p + \frac{1}{3}\tilde\rho}{\tilde p + \tilde\rho}=
\left(1-\frac{8\pi}{3}\tilde\rho r^2 \right)^{h/2}k(t), \label{eq:prep}
\end{equation}
where $k$ is a function of $t$ to be defined from the boundary condition
(\ref{eq:boundary}) or (\ref{eq:boun}).
Thus, (\ref{eq:prep}) and (\ref{eq:boun}) give
\begin{equation}
\tilde\rho=\frac{3m_a}{4\pi a^2},
\end{equation}
\begin{equation}
\tilde p=\frac{\tilde\rho}{3}\Biggl\{\frac{\chi_S (1-2m_a/a)^{1/2} -3\psi_S\xi}
{\psi_S\xi -\chi_S (1-2m_a/a)^{1/2}}\Biggr\}, \label{eq:effepre}
\end{equation}
with
$$
\xi=\left[1-\frac{2m_a}{a}\left(\frac{r}{a}\right)^2\right]^{1/2}
$$
where
\begin{equation}
\chi_S=6(\omega_a^2+1)\frac{m_a}{a}+8\pi a^2 q_a(1+\omega_a)^2,
\end{equation}
and
\begin{equation}\psi_S=2(3\omega_a^2+1)\frac{m_a}{a}+8\pi a^2 q_a(1+\omega_a)^2.
\end{equation}
Using (\ref{eq:m}) and (\ref{eq:nu}) it is easy to obtain expressions
for $m$ and $\nu$:
\begin{equation}
m=m_a(r/a)^3, \label{eq:mass_sch}
\end{equation}
\begin{equation}
e^{\nu}=\Biggl\{\frac{a(\chi_S (1-2m_a/a)^{1/2}-\psi_S\xi)}{4m_a}
\Biggr\}^{2}. \label{eq:nu_sch}
\end{equation}
Thus, the system of equations at the surface can be integrated, but
it is necessary to specify one function of $t$ and the initial data. We choose 
\begin{equation}
L\equiv 4\pi a^2 q_a
\end{equation}
 to be a Gaussian  
\begin{equation}L=L_0 e^{-(t-t_0)^2/\Sigma^2},\end{equation}
with $L_0=M_r/\sqrt{\Sigma \pi}$, $t_0=5.0$ and $\Sigma=0.25$, which corresponds to a pulse
radiating away a fraction of the initial mass $M_r$.
Therefore, the system can be numerically integrated for the following
typical initial conditions:
$$
a(0)=5.0,\,\,m_a(0)=1.0, \,\, \omega_a(0)=0.0.
$$
 The integration was done up to some $t$ guaranteeing well behavior of the physical variables, that is, $\rho >0$; $\rho \ge p$; $|\omega| < 1$; $\omega$, $ q \in \Re$. 
Feeding back the numerical values of $a$, $m_a$ and $\omega_a$ (and their
 derivatives)
 in (\ref{eq:m}) and (\ref{eq:nu}) we obtain $m$ and $\nu$ 
(and their partial derivatives) for any
 value of $r$. Thus, variables $\rho$, $p$, $\omega$ and $q$ can be
 monitored for any piece of the material, via field equations.
 We calculated them for the values $r/a=0.0, \,\, 0.2,\,\, 0.4,\,\, 0.6,\,\, 0.8$ and
 $1.0$. 
 
 We explore a complete range of initial conditions and parameters of integration
 to get physically acceptable results. 
 A radiated mass bigger than $10^{-4}$ and an initially contracting velocity are not permitted. The reason is a complex root calculating the local radial velocity and heat flow for
 some regions of spacetime.
 Representative and acceptable results are shown in Figs. 1--2.
For this model, the energy density and the radius 
of the distribution remain almost constant (within six significant figures).
These features were not reported in the past using radiation coordinates and lead us to the following model. 
Our results clearly show that the heat flow keeps the evolution near quasistaticity (slow evolution). 
Under the same initial compactness used above, that is, $a(0)=5$, we found a possible initial local radial velocity of $\omega_a(0)=-10^{-3}$ and a radiated mass of $M_r=10^{-2}$, producing now an appreciable change in the energy density and the radius of the distribution. These results are shown in Figs. 3--4. For these conditions only the evolution of a bubble is possible
($r/a\approx 0.99\rightarrow1.00$). We do not observe any evidence of thermal peeling \cite{hd97}, that is, positive velocities (expansion) of outer shells and negative velocities (contraction) of the inner shells.
The development of thermal peeling leads to complex roots for the radial
velocity.

We consider now other interior seed model based on the Tolman VI interior solution
\cite{t39}. This model corresponds to a highly compressed Fermi gas. Let us take
\begin{equation}
\tilde\rho=\frac{g}{r^2},
\end{equation}

\begin{equation}
\tilde p = \frac{g[1-9 \alpha (r/a)]}{3[1-\alpha (r/a)] r^2},
\end{equation}
where $g$  and $\alpha$ are functions of $t$, which can be determined using
(\ref{eq:boun}). Thus
\begin{equation}
g=\frac{m_a}{4\pi a} \label{eq:g}
\end{equation}
\begin{equation}
\alpha=\frac{2m_a/a-3\beta}{3[6m_a/a-\beta]}
\end{equation}
\begin{equation}
\beta=2\omega_a^2\frac{m_a}{a}+8\pi a^2 q_a(1+\omega_a)^2.
\end{equation}
Once the metric functions are obtained from (\ref{eq:m}) and (\ref{eq:nu}),
the system of equations at the surface can be again numerically integrated for the following
 initial conditions:
$$
a(0)=8.0,\,\,m_a(0)=1.0, \,\, \omega_a(0)=-0.02.
$$
As before, a radiated mass bigger than $10^{-4}$ is not permitted. But even more, now it is not possible to go inside the distribution without violating real values assumption from the beginning. For that reason the Gaussian has been set to $\Sigma=0.01$ and $t_0=1$. At the surface, see Fig. 5, the results are as expected.

\section{Conclusions}
In this paper we considered heat flow as a transport mechanism in the PQSA. 
Heat flow produces a stable configuration, which is the opposite effect of
viscosity \cite{prrb10}. 
This result indicates that a combination of viscosity (anisotropy) with heat flow may 
be crucial for gravitational collapse or at least just out of equilibrium,
where we expect the PQSA is a good approach. We did additional
tests including anisotropy but its effect is marginal.
For distributions far from equilibrium we find that heat flow is a very 
restrictive transport mechanism.

These results are apparently different for the same configurations in radiation coordinates \cite{allheatflow1}--\cite{allheatflow8}. If the initial distribution is in equilibrium
 the transition from static to postquasistatic, in radiation coordinates, allows the sphere to ``instantaneously'' bypass diffusion stressing. 
But in Schwarzschild coordinates we can follow the transition from the static configuration to the postquasistatic with more resolution. Diffunding radiation strongly interacts with matter. 
As a result, the interior of the distribution is not permitted to go far from equilibrium. 
When including heat flow, there is not PQSA except very close to or at the surface.
 
\acknowledgments  C. P. acknowledges the computing resources provided by the Victorian Partnership for Advanced Computation (VPAC).
 

\begin{thebibliography}{99}
\bibitem{ks79} D. Kazanas and D. Schramm {\it Sources of gravitational
Radiation} (Cambridge: Cambridge University Press, 1979).
\bibitem{hd97} L. Herrera and A. Di Prisco, Phys. Rev. D, {\bf 55}, 2044 (1997).
\bibitem{l88} J. Lattimer, Nucl. Phys. A {\bf 478}, 199 (1988).
\bibitem{prrb10} C. Peralta, L. Rosales, B. Rodr\'\i guez and W. Barreto, 
Phys. Rev. D, {\bf 81}, 104021 (2010).
\bibitem{brm02} W. Barreto, B. Rodr\'\i guez and H. Mart\'\i nez,
Ap. Sp. Sc., {\bf 282}, 581 (2002).
\bibitem{hbds02} L. Herrera, W. Barreto, A. Di Prisco and N. Santos,
 Phys. Rev. D, {\bf 65} 104004 (2002).
\bibitem{b09} W. Barreto Phys. Rev. D {\bf 79}, 10, 107502 (2009).
\bibitem{bcb09} W. Barreto, L. Castillo and E. Barrios  Phys. Rev. D {\bf 80}, 084007 (2009).
\bibitem{bcb10} W. Barreto, L. Castillo and E. Barrios, {\it Bondian frames to couple matter with
radiation}, published on line in General Relativity and Gravitation, 06 March (2010).
\bibitem{hjr80}  L. Herrera, J. Jim\'enez and G. Ruggeri, Phys. Rev.
 D, {\bf 22}, 2305 (1980).
\bibitem{almostall1} L. Herrera and L. N\'u\~nez, Fund. Cosmic Phys.,
 {\bf 14}, 235 (1990).
\bibitem{almostall3} W. Barreto, L. Herrera, and L. N\'u\~nez, Ap. J. {\bf 375}, 663 (1991).
\bibitem{almostall4} W. Barreto, L. Herrera, and N. Santos, Ap. Sp.
Sc. {\bf 187}, 271 (1992).
\bibitem{almostall6} L. Herrera, A. Melfo, L. A. N\'u\~nez, and A. Pati\~no, Ap. J. {\bf 421}, 677 (1994).
\bibitem{almostall7} W. Barreto and A. Da Silva, Gen. Rel. Grav. {\bf 28}, 735 (1996).
\bibitem{allheatflow1} L. Herrera, J. Jim\'enez and M. Esculpi, Phys. Rev. D {\bf 36}, 2986 (1987).
\bibitem{allheatflow2} W. Barreto, L. Herrera and N. Santos. Ap. J. {\bf 344}, 158 
(1989). 
\bibitem{allheatflow6} W. Barreto, L. Herrera and N. Santos, Ap. Sp. Sc. {\bf 187}, 271 (1992). 
\bibitem{allheatflow8} W. Barreto, C. Peralta and L. Rosales, Phys. Rev. D 
{\bf 59}, 024008 (2000). 
\bibitem{m96} J. Mart\'\i nez, Phys. Rev. D, {\bf 53}, 6921 (1996).
\bibitem{b64} H. Bondi, Proc. Royal Soc. London, {\bf A281}, 39 (1964).
\bibitem{t39} R. C. Tolman, Phys. Rev., {\bf 55}, 364 (1939).
\end{thebibliography}
\end{document}